\documentclass[preprint,aps]{revtex4}
\usepackage{graphicx}

\begin{document}

\title{Network topology transition at criticality}

\author{Chung-Pin Chou$^{1}$, Yi-Hua Wang$^{2}$ and Ming-Chiang
Chung$^{3,4}$} \affiliation{$^{1}$Beijing Computational Science
Research Center, Beijing, China} \affiliation{$^{2}$Department of
Physics, National Taiwan Normal University, Taipei 11677, Taiwan}
\affiliation{$^{3}$Department of Physics, National Chung Hsing
University, Taichung 40227, Taiwan} \affiliation{$^{4}$Physics
Division, National Center for Theoretical Science, Hsinchu, 30013,
Taiwan}

\maketitle

%%%%%%%%%%%%%%%%%%%%%%%%%%%%%%%%%%%%%%%%%%%%%%%%%%%%%%%%%%%%%%%%%%%
%                        Abstract                                 %
%%%%%%%%%%%%%%%%%%%%%%%%%%%%%%%%%%%%%%%%%%%%%%%%%%%%%%%%%%%%%%%%%%%

{\bf Many-body systems when continuous phase transition occurs are
mainly built in the interrelationship between particles, implemented
through many-body correlations. Some of them may exhibit so-called
topological order hardly measured by experiments. Therefore we need,
beyond mean-field theory, the complex-systems approach that stresses
the systemic complexity of many-body network at criticality.
According to our previous study \cite{CPCSciRep14}, network space
experiences the homogeneous-heterogeneous transition invisible in
traditional phase transitions. The network robustness can be a
useful indicator to capture the critical phenomena of phase
transitions with/without symmetry breaking. In this work, we
demonstrate the idea of the change of robust networks is
successfully applied to the well-known 1D quantum and 2D classical
XY models.}

%%%%%%%%%%%%%%%%%%%%%%%%%%%%%%%%%%%%%%%%%%%%%%%%%%%%%%%%%%%%%%%%%%%
%                       Introduction                              %
%%%%%%%%%%%%%%%%%%%%%%%%%%%%%%%%%%%%%%%%%%%%%%%%%%%%%%%%%%%%%%%%%%%
Determining the topological order of an interacting quantum system
from its microscopic many-body entanglement is one of the recent
goals of condensed matter theory.
Traditional phases of matter and phase transitions are usually
distinguished by local order parameters.
Consider, for instance, a continuous phase transition, the critical
point is accompanied by a diverging correlation length in Landau's
symmetry breaking framework \cite{SachdevBook99}.
However, it becomes clear that some topologically ordered phases do
not fall into this framework
\cite{WenBook06,NayakRMP08,BernevigBook13}, such as fractional
quantum Hall states \cite{LaughlinPRL83}, quantum spin liquids
\cite{AndersonMRB73} and recently discovered topological insulators
\cite{HasanRMP10,QiRMP11}.
Moreover, characterizing topological phase transitions between them
is a rather difficult task due to the absence of local order
parameters \cite{XGWenIJMP90,CarrBook11}.
There are recent advances in diagnosing the presence of topological
order from knowledge of many-body ground states, including using
entanglement entropy \cite{KitaevPRL06,LevinPRL06}, entanglement
spectrum \cite{LiPRL08,QiPRL12}, modular transformation
\cite{ZhangPRB12}, Chern number \cite{ThoulessPRL82} and $Z_2$
topological invariant \cite{KitaevPU01,KanePRL05}.
They all have brought us closer to being able to tackle such
important questions.

On the other hand, the complex network theory originating from graph
theory has become one standard tool to analyze the structure and
dynamics of real-world systems, which consists of overwhelming
information
\cite{AlbertRMP02,NewmanSIAM03,DorogovtsevAIP02,BoccalettiPR06,DorogovtsevRMP08}.
The building blocks of a complex network include nodes (system's
elements) and links (relation between two elements).
Via the unique patterns of connections, the essential features of a
collection of interacting elements can be unveiled by direct
visualization and network analysis.
Its application is prevailing in many fields, such as sociology,
biology, informatics and many other interdisciplinary studies
\cite{SethnaBook06}.
Thus the generic description is reasonably applicable to condensed
matter systems.

An interesting question in the application of network analysis is:
Can the critical phenomena in condensed matters be observed by
complex network topology \cite{CPCArXiv13}?
The answer could be tracked back to a common phenomenon of many
real-world networks, network robustness, which is characterized by
the connectivity of complex network.
In light of extensive studies in various real-world networks
\cite{AlbertNat99,NewmanPNAS01,DoddsSci03,BackstromProc12}, it will
be interesting to explore the network robustness of many-body
systems.

The focus of this work is to construct the weighted networks for the
one-dimensional (1D) quantum and the two-dimensional (2D) classical
XY models, respectively.
The weight of network links we define carries quantum or classical
correlations between particles in these two models.
A significant observation is that the critical region in the 1D XY
model can be distinguished by the novel network topology, in
addition to typical entanglement entropy.
We find that the entanglement entropy and the network robustness
share similar mathematical structure but with different
interpretations according to graph theory.
The topology arising from the weighted networks is illustrated by
the network robustness.
It is also confirmed that a direct relationship between the change
of the network robustness and the homogeneous-heterogeneous
transition in weight distribution of network links
\cite{AlbertNat00}.
Applying the same network property to the 2D XY model, we show that
as well it may detect the Kosterlitz-Thouless (KT) transition point
obtained by conventional quantities, e.g. spin stiffness.

%%%%%%%%%%%%%%%%%%%%%%%%%%%%%%%%%%%%%%%%%%%%%%%%%%%%%%%%%%%%%%%%%%%
%                          Results                                %
%%%%%%%%%%%%%%%%%%%%%%%%%%%%%%%%%%%%%%%%%%%%%%%%%%%%%%%%%%%%%%%%%%%
\section*{Results}
\textit{1D Quantum XY Model.}
The 1D quantum XY model consists of a chain of $L$ spins with
nearest neighbor interaction and transverse magnetic field
\cite{LatorreArXiv03}.
There is an anisotropy parameter $\gamma$ which is the ratio between
$x$ and $y$ direction interaction, and a tunable transverse magnetic
field strength $\lambda$ identifying criticality in the Hamiltonian,
\begin{equation}
H_{XY}=-\frac{1}{2}\sum_{l}\left(
\frac{1+\gamma}{2}\sigma_{l}^{x}\sigma_{l+1}^{x}+\frac{1-\gamma}{2}\sigma_{l}^{y}\sigma_{l+1}^{y}+\lambda\sigma_{l}{z}
\right),
\end{equation}
where $l$ labels the spins and $\sigma_{u}^{l}(u=x,y,z)$ are the
Pauli matrix.
In the spin-$\frac{1}{2}$ chain, there are
ferromagnetic-paramagnetic transitions.
The phase transition is based on the transverse magnetic field and
the anisotropy parameter.
Let the anisotropic parameter $\gamma=0$, we recover the XX
Hamiltonian with transverse magnetic field,
\begin{equation}
H_{XX}=-\frac{1}{2} \sum_{l}\left(
\frac{1}{2}\sigma_{l}^{x}\sigma_{l+1}^{x}+\frac{1}{2}\sigma_{l}^{y}\sigma_{l+1}^{y}+\lambda\sigma_{l}{z}
\right).\label{XXM}
\end{equation}
By applying Jordan-Wigner transformation to Eq.(\ref{XXM}), we
derive the Majorana fermionic Hamiltonian which is skew-symmetric
\cite{KitaevAP03,LevinPRB05}.
Further using Fourier transformation and Bogoliubov transformation,
the ground state is expressed with the correlation matrix
$\Gamma_{mn}^{A}$ of the Majorana operator $a_n$
\begin{equation}
a_ma_n=\delta_{m,n}+i\Gamma_{mn}^{A}
\end{equation}
where
\begin{eqnarray}
\Gamma^{A}=\left(
\begin{array}{cccc}
\Pi_0  &\Pi_1 &\ldots&\Pi_{L-1}\\
-\Pi_1 &\Pi_0 &     &\vdots  \\
 \vdots &     &\ddots     &\vdots   \\
-\Pi_{L-1}&\ldots &\ldots&\Pi_{0} \\
\end{array}\right),\\
\Pi_l=\left(
\begin{array}{cc}
0&g_l\\
-g_l&0\\
\end{array}\right),
\end{eqnarray}
with real coefficients $g_l$ given, in the limit of an infinite
chain, by
\begin{equation}
g_l=\frac{1}{2\pi}\int_{0}^{2\pi}d\phi e^{il\phi}
\frac{cos{\phi}-\lambda-i\gamma
sin{\phi}}{|cos{\phi}-\lambda-i\gamma sin{\phi}|}.
\end{equation}
The adjacency matrix $\hat{A}_{ij}$ in the network representation is
defined by the correlation matrix $\Gamma^A$.
Note that the size of network space would be $2L$ owing to two
Majorana fermions at each site.
The weight of the network link is just the correlation matrix
element between two nodes.

It is well-known that there exists the critical region
$(0<\lambda<1)$ in the XX Hamiltonian with transverse magnetic
field.
As increasing $\lambda$, the paramagnetic phase would pass through
the critical point ($\lambda$=1) and change into the ferromagnetic
phase.
For the critical XX model, the entanglement entropy of the region
$\Omega=\left\{1,L\right\}$ scales as
\begin{equation}
S=\frac{1}{3}\log_{2}L+O(1).
\end{equation}
The maximum entropy is reached when $\lambda=0$.
The entropy decreases while the transverse magnetic field increases
until $\lambda=1$ when the system goes into the ferromagnetic phase.
Similar universal behavior of the model can be proven for Renyi
entropy of order $n$,
\begin{equation}
S_n=\frac{1}{1-n}\log\Big(Tr(\rho^{n})\Big),\label{renyi}
\end{equation}
of which the reduced density matrix can be reconstructed from the
correlation matrix $\Gamma^A$. On the other hand, the network
robustness (defined in Section Methods) can be derived from
\begin{equation}
R=ln\Big(\frac{1}{N}\sum_{n=1}^{\infty}\frac{n!}{N_n}\Big)
\end{equation}
where $N_n=Tr\big(\hat{A}^n\big)$ represents the strength of loops
of length $n$ in network space.
At first glance, interestingly, the network robustness has the
similar mathematical structure as the Renyi entropy when the
adjacency matrix is defined as the reduced density matrix.
The Renyi entropy of order $n$ is just the $n$-th term of the sum in
Eq.(\ref{renyi}).
In other words, the Renyi entropy of order $n$ partly describes the
robustness of abstract network space.
We have to consider the Renyi entropy of all orders if the entire
network space needs to be explored.
Therefore, we reasonably speculate that the critical behavior of the
XX model is more properly described by the network robustness.

\begin{figure*}
\centering
\includegraphics[width=0.6\textwidth]{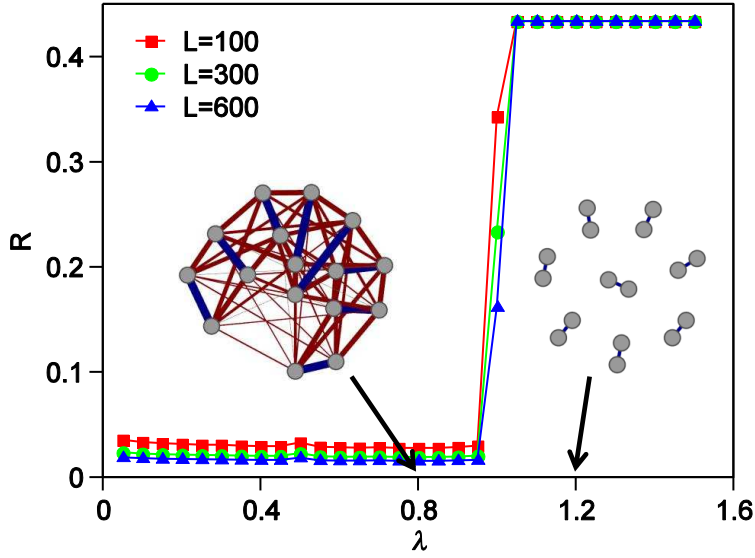}
\caption{\label{Figure 1}The network robustness $R$ of the 1D
quantum XY model vs transverse magnetic field strength $\lambda$ for
different chain length $L$. Network representations for $\lambda$
below and above the critical point are shown in the inset.}
\end{figure*}

In Fig.\ref{Figure 1}, we calculate the network robustness of the
critical XX model as a function of transverse magnetic field
strength.
Within the critical region $(0<\lambda<1)$, the complex network
system is rather fragile to random failures.
But after being thrust into the ferromagnetic phase $(\lambda>1)$,
the robustness of the network recovers.
The interesting phenomenon can be understood by looking into the
network space in details.
Taking an example of $\lambda=0.8$, the shape of the network whose
nodes connect to each other with different link weights looks much
like a "dimer liquid".
Here the dimer we call is the strongest link highlighted in blue
color in the inset of Fig.\ref{Figure 1}.
In fact, the dimer is formed by two Majorana fermions at the same
site in real space.
The other weak links highlighted in red color attempt to connect the
dimers to form the "liquid-like" pattern.
Above $\lambda=1.0$, most of links between dimers vanish, and the
network structure turns a "dimer liquid" into a "dimer gas".
In the ferromagnetic region, all dimers become independent with
respect to each other leading to the appearance of the dimer gas in
the network space.
It seems to involve some sort of "liquid-gas" transition in the
network pattern.
Owing to heterogeneous distribution of link weights (explained
below), there exists the more robust network structure in the
ferromagnetic phase.
Furthermore, the critical point $\lambda=1$ can be clearly found by
the scenario as well.

\begin{figure*}
\centering
\includegraphics[width=0.8\textwidth]{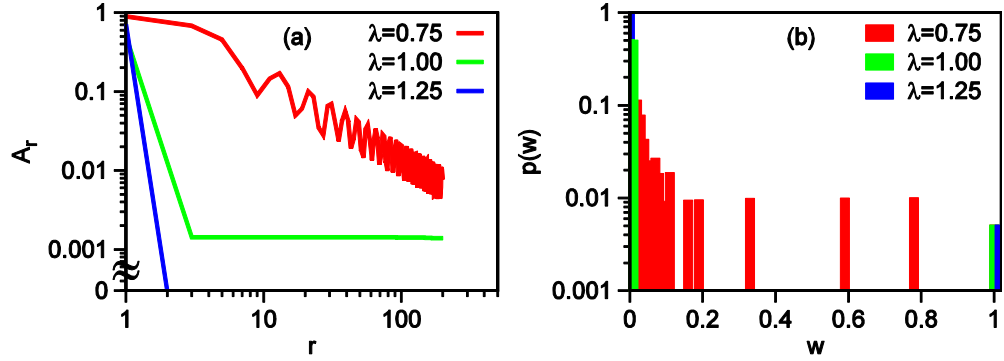}
\caption{\label{Figure 2}(a) The adjacency matrix element $A_{r}$ as
a function of distance $r$ and (b) The probability distribution
$p(w)$ of weight $w$ of network links for different transverse
magnetic field strength $\lambda$. The bin size is chosen for clear
demonstrations. The chain length $L$=200.}
\end{figure*}

In Fig.\ref{Figure 2}, we analyze the homogeneous and heterogeneous
network topologies in the 1D XY model.
It shows in Fig.\ref{Figure 2}(a) that the distance $r(=|i-j|)$
dependence of the adjacency matrix element $\hat{A}_{r}$ changes as
the phase transition occurs at $\lambda=1$.
In the critical region $(\lambda<1)$, the critical phase shows
power-law behavior at all distances.
Above the critical point $(\lambda>1)$, the ferromagnetic phase has
only the non-zero matrix elements at $r=1$ leading to the extremely
heterogeneous weight distribution of network links.
As shown in Fig.\ref{Figure 2}(b), the evolution of the probability
distribution $p(w)$ of weight $w$ of network links can provide a
clear picture to illustrate the change of network topologies.
First, the weights of network links continuously distribute in the
network space when $\lambda<1$.
As increasing $\lambda$, the weight distribution becomes more
heterogeneous.
Above $\lambda=1$, it even comes to two delta functions at $w=0$ and
$1$ leading to extreme heterogeneity.
It is a well-known fact that complex weighted networks with the
heterogeneous weight distribution of network links are much more
robust to random failures or attacks \cite{AlbertNat00,WangCM05}.
To put it another way, the homogeneous weight distribution is one of
important features of fragile networks.
Therefore, it can be expected that there is a
homogeneous-heterogeneous network transition hidden in the quantum
phase transition as changing network robustness.
It may allow us to further define the network robustness as a
detector in network space for identifying the quantum phase
transition.

\textit{2D Classical XY Model.} The other example is the 2D
classical XY model in a square lattice of size $N$ described by
\cite{KosterlitzJPC73,KosterlitzJPC74}
\begin{equation}
-\sum_{\left\langle i,j\right\rangle}\vec{S}_i\cdot
\vec{S}_j=\sum_{\left\langle i,j
\right\rangle}\cos\left(\theta_i-\theta_j\right),
\end{equation}
where $\theta_i$ is the angle of the 2D spin vector $\vec{S}_i$  at
site $i$.
In contrast to 1D quantum XY system, the true long-range order is
completely washed out by thermal or quantum fluctuations and only
its topology remains.
In fact, its low-temperature phase forms a quasi-long-range order
originating from the power-law spin correlation decay.
There is a phase transition from this phase to the high-temperature
disordered phase whose correlations decay exponentially with
distances.
Such a transition is known as the KT transition associated with the
disappearance of the quasi-long-range order.
Near the topological phase transition the system begins to lose spin
stiffness that shows up a universal jump near the transition
temperature $T_c=0.9$.
This approach has been often used to numerically extract $T_c$
\cite{NelsonPRL77,WeberPRB87,HasenbuschJSM08}.
Other than the 1D quantum case, we define the adjacency matrix as
the spin-spin correlation function:
\begin{equation}
\hat{A}_{ij}\equiv|\vec{S}_i\cdot\vec{S}_j|,
\end{equation}
which can be calculated by using the standard Monte Carlo
simulation.
In what follows, we will show that the network topology also enables
the network robustness to illustrate the KT transition.

\begin{figure*}
\centering
\includegraphics[width=0.6\textwidth]{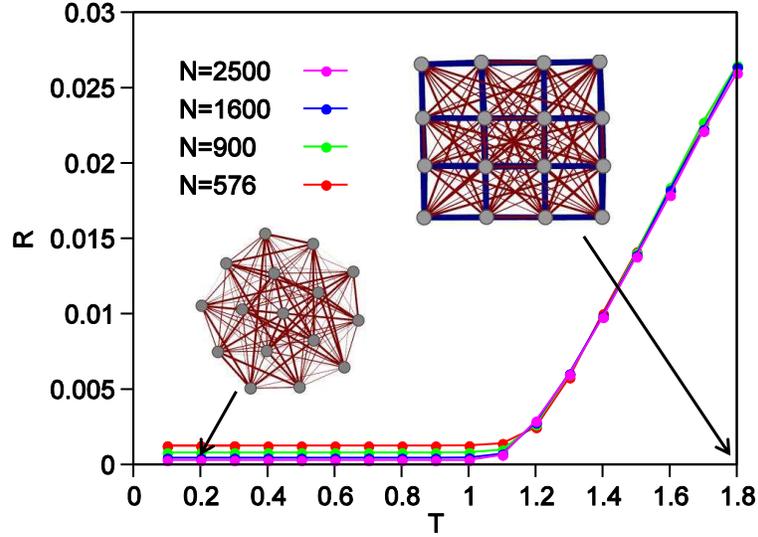}
\caption{\label{Figure 3}The network robustness $R$ of the 2D
classical XY model vs temperature $T$ for different lattice size
$N$. Network representations for $T$ below and above the critical
point are shown in the inset.}
\end{figure*}

Figure \ref{Figure 3} shows the temperature evolution of the network
robustness in the 2D XY model.
As mentioned above, the low-temperature topology gives rise to the
power-law correlation decay, and hence links and nodes are
homogeneously arranged in an irregular network pattern as
illustrated in the inset.
The network representation also looks like a "liquid" pattern.
When temperature approaches $T_c$, the short-range spin correlations
become more and more dominant.
At much higher temperature $(T=1.8)$, each node has four strongest
links to its neighbors as shown in the inset.
The network topology is equivalent to a torus which corresponds to a
square lattice with periodic boundary conditions.
Now the network pattern becomes "crystal-like" in the network space.
Hence, a "liquid-crystal" transition at $T_c$ can be expected to
happen in the network space.
The reason behind the topology is that extremely short-range spin
correlations of the high-temperature disordered phase indicate much
larger link weights between the nearest-neighboring sites, and thus
the network structure resembles a torus.
Conventionally, spin stiffness has usually been used to determine
the transition temperature.
Here we demonstrate that the network robustness also have the
ability to detect the KT transition.

\begin{figure*}
\centering
\includegraphics[width=0.8\textwidth]{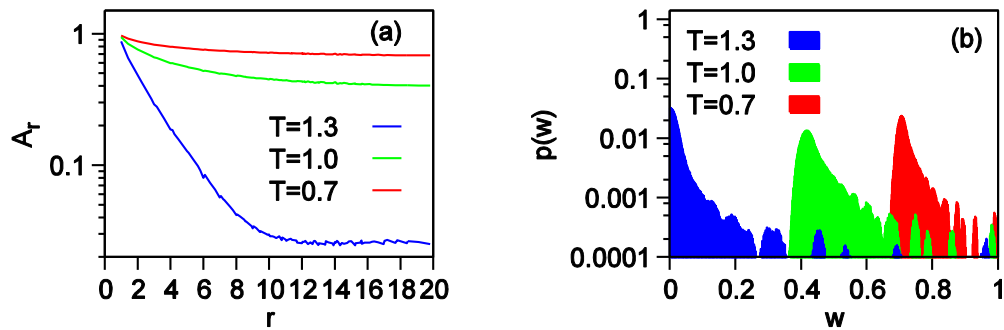}
\caption{\label{Figure 4}(a) The adjacency matrix element $A_{r}$ as
a function of distance $r$ and (b) The probability distribution
$p(w)$ of weight $w$ of network links for different temperature $T$.
The lattice size $N$=900.}
\end{figure*}

Consider translational invariance, Figure \ref{Figure 4}(a) presents
how the adjacency matrix element $\hat{A}_{r}$ changes as the KT
transition occurs.
For low temperature $(T=0.7)$, the quasi-long-range-ordered phase
has the power-law correlation at large distances leading to the
"liquid-like" network pattern.
Not much far from the 1D example, Figure \ref{Figure 4}(b) shows
that the weight distribution of network links looks log-normal.
In other words, most links center their weights around $w=0.75$ and
display homogeneous distribution with a short tail ending at $w=1$.
The heterogeneity of the weight distribution appears as approaching
the critical point $(T_{c}\sim0.9)$.
Above the critical point $(T=1.3)$, the high-temperature phase
instead shows that the spin-spin correlation is exponentially
decaying with distances.
The exponentially decaying correlation function in real space
results in the "crystal-like" structure in network space.
In the meantime, the peak of the distribution is moved to $w\sim0$
with a long tail so that it becomes much more heterogeneous.
Due to much broader weight distributions than the 1D case, it is
rather difficult to come up with a clear order parameter in the 2D
classical XY model.
The same reasoning from the network topology can be applied to other
many-body systems without local order parameters.
Even so, the network robustness can still be considered as a useful
quantity to characterize the KT transition, in addition to
traditional spin stiffness.

\textit{Conclusion.} We have introduced the novel complex network
analysis for helping us find how the network topology changes as a
system undergoes the phase transition.
In the 1D quantum XY model, the topology of the complex network is
transited from "dimer liquid" to "dimer gas".
At the critical point, a homogeneous-heterogeneous transition occurs
in the weight distribution of network links.
In contrast to entanglement entropy used in many literatures, the
network robustness including more information about connectivity of
the complex network may provide a more complete map to uncover the
universal properties of the quantum phase transition.
Thus the network robustness can be fairly considered as another
detector in the quantum XY model.

In the 2D classical XY model, we have found that the critical
phenomena can also be described by the change of the weighted
network topology.
The similar behavior that the structure of the weighted network
varies from homogeneity to heterogeneity brings about the appearance
of the network robustness to random failures.
The robustness of the complex network is able to uncover a wealth of
topological information underneath the classical spin-spin
correlation, and further comprehend the mechanism of the phase
transition without local order parameters.
Notice that the complex network analysis is absolutely not designed
for effectively calculating correlation functions but principally
extracting the universal properties of the phase transitions without
any symmetry breaking.
The picture behind the weight distribution of network links can even
provide significant information to comprehend the generic phase
transitions no matter symmetry is broken or not.
Therefore we propose that the network analysis approach presented in
this work gives more general recipe for studying a variety of phase
transitions in condensed matters.

\section*{Methods}

In the language of complex network, each network of $N$ nodes is
described by its $N\times N$ adjacency matrix representation
$\hat{A}$ \cite{AlbertRMP02}.
A number of real systems, e.g. transportation networks, neural
networks and so on, are better captured by the weighted network in
which the link use weight to quantify their strength.
In this work, we consider lattice sites as nodes of the weighted
network, with each weighted link between nodes $i$ and $j$ expressed
by the element of the adjacency matrix $\hat{A}_{ij}$.
The link carries the weight containing detailed information about
the correlation between particles at different sites.
In other words, the nodes (or say, lattice sites) are not connected
by lattice bond but correlation between particles at corresponding
sites so that a complete weighted network is formed.

The symmetric adjacency matrix $\hat{A}_{ij}$ in these two examples
only has real entries.
Hence the network gives rise to the signed structure whose link
weights are allowed to be either positive or negative.
A similar example is an acquaintance network in which we denote
friendship by a positive link and animosity by a negative link
\cite{FacchettiPNAS11}.
Further definition of the two adjacency matrices corresponding to
positive and negative correlations between lattice sites is
possible.
However, we find the results would not be changed if disregarding
the sign of $\hat{A}_{ij}$, because the negative part is a tiny
minority of the entire matrix.
For simplicity, we will take the absolute value of $\hat{A}_{ij}$.
Following the convention in the weighted complex network
\cite{SaramakiPRE07}, we also take the absolute value of
$\hat{A}_{ij}$ normalized by the maximum weight in the network
$\frac{|A_{ij}|}{\max A_{ij}}$.

In complex network theory, a variety of network measures have been
proposed to detect the structural robustness
\cite{HararyNet83,KrisCMA87,EsfIPL88,BauerCC90}.
Natural connectivity derived mathematically from the graph spectrum
has been introduced to measure the robustness of the weighted
network structure \cite{BarahonaCPL10,BarahonaChao12,EstradaPR12}.
More precisely, the network robustness can be described by the
ability of a network to continue performing well after removing a
fraction of nodes or links.
One possible definition of the network robustness is similar to
natural connectivity, given by
\begin{equation}
R=ln\left(\frac{1}{N}\sum_{i=1}^{N}e^{\overline{\lambda}_{i}}\right)
\end{equation}
where
$\overline{\lambda}_{i}(\equiv\frac{\lambda_{i}}{\max\lambda_{i}})$
stands for the normalized eigenvalue of the adjacency matrix in
order to avoid the enhancement of the network robustness as
increasing the number of nodes $N$.
Based on the definition of the network robustness, the reasoning
behind the homogeneous-heterogeneous network transition is now
clear.
With the network transition from homogeneity to heterogeneity,
exponentially decaying many-body correlation between nodes resulting
far away from the critical point is the root cause of the robustness
of complex network.
The network robustness thus provides more information to comprehend
traditional phase transitions than mean-field picture.

\vspace{3mm}

\noindent {\bf Acknowledgement}\\
This work is supported by CAEP and MST.

\vspace{3mm}

\noindent{\bf Competing Financial Interests:}\\
The authors declare no competing financial interests.

\vspace{3mm}

\noindent {\bf Correspondence}\\
Correspondence and requests for materials should be addressed to
cpc63078@gmail.com.

\newpage

\end{document}